\documentclass{article}
\usepackage{ismir,amsmath,cite,url}
\usepackage{graphicx}
\usepackage{color}
\usepackage{subfig}
\usepackage{booktabs}
\usepackage{mathtools}
\usepackage{enumitem}
\usepackage{siunitx}
\usepackage[capitalise]{cleveref}

\creflabelformat{equation}{#2#1#3}

\title{The NES Music Database: A multi-instrumental dataset with expressive performance attributes}

\threeauthors
  {Chris Donahue} {UC San Diego \\ {\tt cdonahue@ucsd.edu}}
  {Huanru Henry Mao} {UC San Diego \\ {\tt hhmao@ucsd.edu}}
  {Julian McAuley} {UC San Diego \\ {\tt jmcauley@ucsd.edu}}

\newcommand{\nesmdb}{NES-MDB}

\begin{document}

\maketitle
\begin{abstract}
Existing research on music generation focuses on composition, 
but often ignores the expressive performance characteristics required for plausible renditions of resultant pieces. 
In this paper, 
we introduce the Nintendo Entertainment System Music Database (\nesmdb), 
a large corpus allowing for separate examination of the tasks of composition and performance. 
\nesmdb{} contains thousands of multi-instrumental songs composed for playback by the compositionally-constrained NES audio synthesizer. 
For each song, 
the dataset contains a musical score for four instrument voices 
as well as expressive attributes for the dynamics and timbre of each voice. 
Unlike datasets comprised of General MIDI files, 
\nesmdb{} includes all of the information needed to render \emph{exact} acoustic performances of the original compositions. 
Alongside the dataset, 
we provide a tool that renders generated compositions as NES-style audio by emulating the device's audio processor. 
Additionally, 
we establish baselines for the tasks of composition, 
which consists of learning the semantics of composing for the NES synthesizer, 
and performance, 
which involves finding a mapping between a composition and realistic expressive attributes. 
\end{abstract}

\section{Introduction}\label{sec:introduction}


The problem of automating music composition is a challenging pursuit with the potential for substantial cultural impact. 
While 
early systems were hand-crafted by musicians to encode musical rules and structure~\cite{nierhaus2009algorithmic}, 
recent attempts view composition as a statistical modeling problem using machine learning~\cite{briot2017deep}. 
A major challenge to casting this problem in terms of modern machine learning methods is building representative datasets for training.
So far, 
most datasets only contain information necessary to model the semantics of music composition, 
and lack details about how to translate these pieces into nuanced performances. 
As a result, 
demonstrations of machine learning systems trained on these datasets sound rigid and deadpan. 
The datasets that do contain expressive performance characteristics 
predominantly focus on solo piano~\cite{poliner2006discriminative,sapp2007comparative,flossmann2010magaloff} rather than multi-instrumental music. 

A promising source of multi-instrumental music that contains both compositional and expressive characteristics is music from early videogames. 
There are nearly $1400$\footnote{Including games released only on the Japanese version of the console} 
unique games licensed for the Nintendo Entertainment System (NES),
all of which include a musical soundtrack. 
The technical constraints of the system's audio processing unit (APU) impose a maximum of four simultaneous monophonic instruments. 
The machine code for the games preserves the \emph{exact} expressive characteristics needed to perform each piece of music as intended by the composer. 
All of the music was composed in a limited time period and, as a result, is more stylistically cohesive than other large datasets of multi-instrumental music. 
Moreover, 
NES music is celebrated by enthusiasts 
who continue to listen to and compose music for the system~\cite{collins2008game}, 
appreciating the creativity that arises from resource limitations.

In this work, 
we introduce \nesmdb, 
and formalize two primary tasks for which the dataset serves as a large test bed.
The first task consists of learning the semantics of composition on a \emph{separated score}, 
where individual instrument voices are explicitly represented. 
This is in contrast to the common \emph{blended score} approach for modeling polyphonic music, 
which examines reductions of full scores. 
The second task 
consists of mapping compositions onto sets of expressive performance characteristics. 
Combining strategies for separated composition and expressive performance yields an effective pipeline for generating NES music \emph{de novo}. 
We establish baseline results and reproducible evaluation methodology for both tasks. 
A further contribution of this work is a  
library that converts between NES machine code (allowing for realistic playback) and representations suitable for machine learning.\footnote{\url{https://github.com/chrisdonahue/nesmdb}} 



\section{Background and task descriptions}

\begin{figure}[t]
\subfloat[Blended score (degenerate)]{
	\label{fig:rep_blend}
    \includegraphics[width=0.98\linewidth]{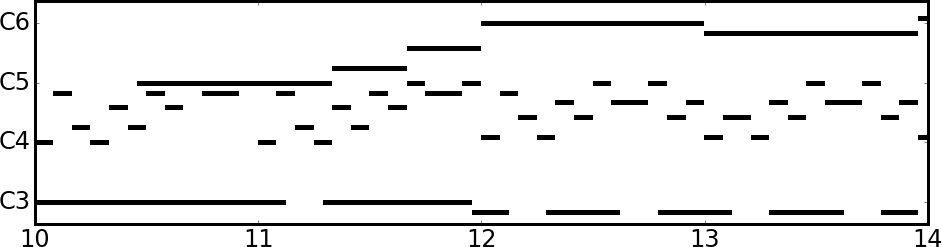}
} \\
\subfloat[Separated score (melodic voices \textbf{top}, percussive voice \textbf{bottom})]{
	\label{fig:rep_sep}
    \includegraphics[width=0.98\linewidth]{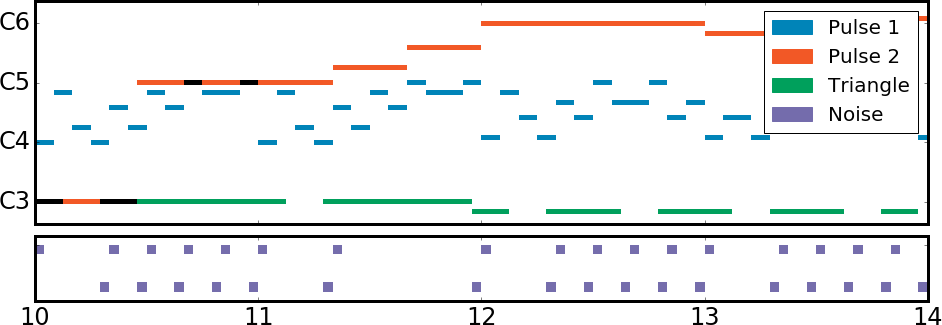}
} \\
\subfloat[Expressive score (includes dynamics and timbral changes)]{
	\label{fig:rep_perf}
    \includegraphics[width=0.98\linewidth]{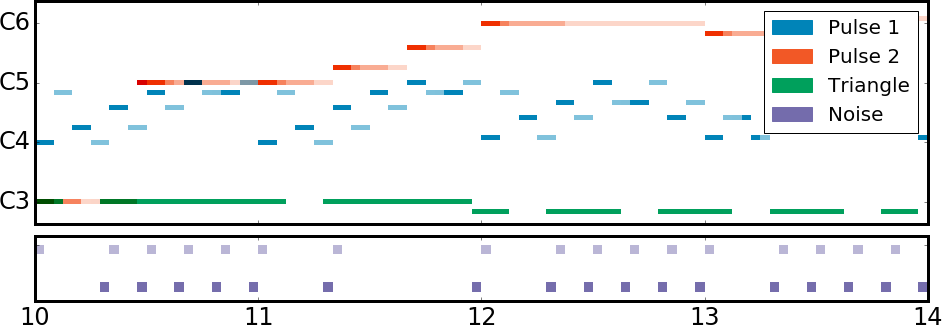}
}
\caption{Three representations (rendered as piano rolls) for a segment of \emph{Ending Theme} from \emph{Abadox} (1989) by composer Kiyohiro Sada. The blended score (\cref{fig:rep_blend}), used in prior polyphonic composition research, is degenerate when multiple voices play the same note.}
\label{globallable}
\end{figure}

Statistical modeling of music seeks to learn the distribution $P(\text{music})$ from human compositions $\bm{c}~\sim~P(\text{music})$ in a dataset $\mathcal{M}$. 
If this distribution could be estimated accurately,
a new piece could be composed simply by sampling. 
Since the space of potential compositions is exponentially large, 
to make sampling tractable, 
one usually assumes a factorized distribution. 
For \emph{monophonic} sequences, 
which consist of no more than one note at a time, 
the probability of a sequence $\bm{c}$ (length $T$) might be factorized as
\begin{equation}
\label{eq:mono}
P(\bm{c}) = P(n_1) \cdot P(n_2 \mid n_1) \cdot \ldots \cdot P(n_{T} \mid n_{t < T}).
\end{equation}

\subsection{Blended composition}


While \cref{eq:mono} may be appropriate for modeling compositions for monophonic instruments, 
in this work we are interested in the problem of multi-instrumental \emph{polyphonic} composition, 
where multiple monophonic instrument \emph{voices} may be sounding simultaneously. 
Much of the prior research on this topic~\cite{boulanger2012modeling,chung2014empirical,johnson2017generating} represents music in a blended score representation.
A blended score $B$ 
is a sparse binary matrix of size $N \times T$, where $N$ is the number of possible note values, 
and $B[n, t] = 1$ if any voice is playing note $n$ at timestep $t$ or $0$ otherwise 
(\cref{fig:rep_blend}). 
Often, $N$ is constrained to the $88$ keys on a piano keyboard, 
and $T$ is determined by some subdivision of the meter, such as sixteenth notes. 
When polyphonic composition $\bm{c}$ is represented by $B$, 
statistical models often factorize the distribution as a sequence of \emph{chords}, the columns $B_t$:
\begin{equation}
\label{eq:blend}
P(\bm{c}) = P(B_1) \cdot P(B_2 \mid B_1) \cdot \ldots \cdot P(B_T \mid B_{t < T}).
\end{equation}

This representation simplifies the probabilistic framework of the task, 
but it is problematic for music with multiple instruments (such as the music in \nesmdb). 
Resultant systems must provide an additional mechanism for assigning notes of a blended score to instrument voices, 
or otherwise render the music on polyphonic instruments such as the piano.

\subsection{Separated composition}

Given the shortcomings of the blended score, 
we might prefer models which operate on a separated score representation (\cref{fig:rep_sep}). 
A separated score $S$ is a matrix of size $V \times T$, where $V$ is the number of instrument voices, 
and $S[v, t] = n$, the note $n$ played by voice $v$ at timestep $t$. 
In other words, the format encodes a monophonic sequence for each instrument voice. 
Statistical approaches to this representation can explicitly model the relationships between various instrument voices by
\begin{equation}
\label{eq:sep}
P(\bm{c}) = \prod_{t=1}^{T} \prod_{v=1}^{V} P(S_{v,t} \mid S_{v,\hat{t} \neq t}, S_{\hat{v} \neq v, \forall \hat{t}})
.
\end{equation}

This formulation explicitly models the dependencies between $S_{v, t}$, 
voice $v$ at time $t$, 
and every other note in the score. 
For this reason, 
\cref{eq:sep} more closely resembles the process by which human composers write multi-instrumental music, 
incorporating temporal and contrapuntal information. 
Another benefit is that resultant models can be used to harmonize with existing musical material, 
adding voices conditioned on existing ones. 
However, 
any non-trivial amount of temporal context introduces high-dimensional interdependencies,
meaning that
such a formulation would be challenging to sample from. 
As a consequence, 
solutions
are often restricted to only take past temporal context into account, 
allowing for simple and efficient ancestral sampling 
(though Gibbs sampling can also be used to 
sample from \cref{eq:sep}~\cite{hadjeres2017deepbach,huang2016counterpoint}).

Most existing datasets of multi-instrumental music  
have uninhibited polyphony, 
causing a separated score representation to be inappropriate. 
However, 
the hardware constraints of the NES APU impose a strict limit on the number of voices, 
making the format ideal for \nesmdb.

\subsection{Expressive performance}

Given a piece of a music, 
a skilled performer will embellish the piece with \emph{expressive characteristics}, 
altering the timing and dynamics to deliver a compelling rendition. 
While a few instruments have been augmented to capture this type of information symbolically (e.g.~a Disklavier), 
it is rarely available for examination in datasets of multi-instrumental music. 
Because NES music is comprised of instructions that recreate an exact rendition of each piece, 
expressive characteristics controlling the velocity and timbre of each voice are available in \nesmdb{} (details in \cref{sec:expr}). 
Thus, each piece can be represented as an \emph{expressive score} (\cref{fig:rep_perf}), 
the union of its separated score and expressive characteristics.

We consider the task of mapping a composition $\bm{c}$ onto expressive characteristics $\bm{e}$. Hence, we would like to model $P(\bm{e} \mid \bm{c})$, 
and the probability of a piece of music $P(\bm{m})$ 
can be expressed as $P(\bm{e} \mid \bm{c}) \cdot P(\bm{c})$, 
where $P(\bm{c})$ is from \cref{eq:sep}. 
This allows for a convenient pipeline for music generation 
where a piece of music is first composed with binary amplitudes 
and then mapped to realistic dynamics, as if interpreted by a performer.

\subsection{Task summary}
\label{sec:task_summary}

In summary, 
we propose three tasks for which \nesmdb{} serves as a large test bed. 
A pairing of two models that address the second and third tasks can be used to generate novel NES music.


\begin{enumerate}
\item The \emph{blended composition} task (\cref{eq:blend}) models the semantics of blended scores (\cref{fig:rep_blend}). This task is more useful for benchmarking new algorithms than for NES composition.
\item The \emph{separated composition} task consists of modeling the semantics of separated scores (\cref{fig:rep_sep}) using the factorization from \cref{eq:sep}.
\item The \emph{expressive performance} task seeks to map separated scores to expressive characteristics needed to generate an expressive score (\cref{fig:rep_perf}). 
\end{enumerate}

\section{Dataset Description}
\label{sec:dataset}

The NES APU consists of five monophonic instruments: 
two pulse wave generators (P1/P2), 
a triangle wave generator (TR), 
a noise generator (NO), 
and a sampler which allows for playback of audio waveforms stored in memory. 
Because the sampler may be used to play melodic or percussive sounds, 
its usage is compositionally ambiguous and we exclude it from our dataset. 

\begin{table}[t]
\centering
\begin{tabular}{l|r}
	\toprule
    \# Games & $397$ \\ 
    \# Composers & $296$ \\ 
    \# Songs & $5,278$ \\
    \# Songs w/ length $>10s$ & $3,513$ \\
    \# Notes & $2,325,636$ \\
    Dataset length & $46.1$ hours \\
    $P(\text{Pulse 1 On})$ & $0.861$ \\
    $P(\text{Pulse 2 On})$ & $0.838$ \\
    $P(\text{Triangle On})$ & $0.701$ \\
    $P(\text{Noise On})$ & $0.390$ \\
    Average polyphony & $2.789$ \\
    \bottomrule
\end{tabular}
\caption{Basic dataset information for \nesmdb.}
\label{tab:stats}
\end{table}

In raw form, 
music for NES games exists as machine code living in the read-only memory of cartridges, 
entangled with the rest of the game logic. 
An effective method for extracting a musical transcript is to emulate the game and log the timing and values of writes to the APU registers. 
The video game music (VGM) format\footnote{\url{http://vgmrips.net/wiki/VGM_Specification}} was designed for precisely this purpose, and 
consists of an ordered list of writes to APU registers with \SI{44.1}{\kilo\hertz} timing resolution. 
An online repository\footnote{\url{http://vgmrips.net/packs/chip/nes-apu}} contains over $400$ NES games logged in this format. 
After removing duplicates, 
we split these games into distinct training, validation and test subsets with an $8$:$1$:$1$ ratio, 
ensuring that no composer appears in two of the subsets. 
Basic statistics of the dataset appear in \cref{tab:stats}. 

\subsection{Extracting expressive scores}
\label{sec:expr}

Given the VGM files, 
we emulate the functionality of the APU to yield an expressive score (\cref{fig:rep_perf}) 
at a temporal discretization of \SI{44.1}{\kilo\hertz}. 
This rate is unnecessarily high for symbolic music, 
so we subsequently downsample the scores.\footnote{We also release \nesmdb{} in MIDI format with no downsampling} 
Because the music has no explicit tempo markings, 
we accommodate a variety of implicit tempos by choosing a permissive downsampling rate of \SI{24}{\hertz}.
By removing dynamics, timbre, and voicing at each timestep, we derive separated score (\cref{fig:rep_sep}) and blended score (\cref{fig:rep_blend}) versions of the dataset. 

\begin{table}[h]
\centering
\begin{tabular}{c|ccc}
	\toprule
    Instrument & Note & Velocity & Timbre \\
    \midrule
    Pulse $1$ (P1) & $\{0, 32, \ldots, 108\}$ & $[0, 15]$ & $[0, 3]$ \\
    Pulse $2$ (P2) & $\{0, 32, \ldots, 108\}$ & $[0, 15]$ & $[0, 3]$ \\
    Triangle  (TR) & $\{0, 21, \ldots, 108\}$ &   & \\
    Noise     (NO) & $\{0, 1, \ldots, 16\}$ & $[0, 15]$ & $[0, 1]$ \\
    \bottomrule
\end{tabular}
\caption{Dimensionality for each timestep of the expressive score representation (\cref{fig:rep_perf}) in \nesmdb.}
\label{tab:instruments}
\end{table}

In \cref{tab:instruments}, 
we show the dimensionality of the instrument states at each timestep of an expressive score in \nesmdb. 
We constrain the frequency ranges of the \emph{melodic} voices (pulse and triangle generators) 
to the MIDI notes on an $88$-key piano keyboard ($21$ through $108$ inclusive, though the pulse generators cannot produce pitches below MIDI note $32$). 
The \emph{percussive} noise voice has $16$ possible ``notes'' 
(these do not correspond to MIDI note numbers) 
where higher values have more high-frequency noise. 
For all instruments, 
a note value of $0$ indicates that the instrument is not sounding (and the corresponding velocity will be $0$). 
When sounding, the pulse and noise generators have $15$ non-linear velocity values, 
while the triangle generator has no velocity control beyond on or off. 

Additionally, 
the pulse wave generators have $4$ possible duty cycles (affecting timbre), 
and the noise generator has a rarely-used mode where it instead produces metallic tones. 
Unlike for velocity, 
a timbre value of $0$ corresponds to an actual timbre setting and does not indicate that an instrument is muted. 
In total, the pulse, triangle and noise generators have state spaces of sizes $4621$, $89$, and $481$ respectively---around $40$ bits of information per timestep for the full ensemble. 



\section{Experiments and discussion}


Below, we describe our evaluation criteria for experiments in separated composition and expressive performance. 
We present these results only as statistical baselines for comparison; 
results do not necessarily reflect a model's ability to generate compelling musical examples.\\[2pt]
\noindent\textbf{Negative log-likelihood and Accuracy}
Negative log-likelihood (NLL) is the (log of the) likelihood that a model assigns to unseen real data (as per \cref{eq:sep}). 
A low NLL averaged across unseen data may indicate that a model captures semantics of the data distribution.  
Accuracy is defined as the proportion of timesteps where a model's prediction is equal to the actual composition. 
We report both measures for each voice, as well as aggregations across all voices by summing (for NLL) and averaging (for accuracy).\\[2pt]
\noindent\textbf{Points of Interest (POI).}
Unlike other datasets of symbolic music, 
\nesmdb{} is temporally-discretized at a high, fixed rate (\SI{24}{\hertz}), 
rather than at a variable rate depending on the tempo of the music. 
As a consequence, 
any given voice has around an $83\%$ chance of playing the same note as that voice at the previous timestep. 
Accordingly, 
our primary evaluation criteria focuses on musically-salient \emph{points of interest} (POIs), 
timesteps at which a voice deviates from the previous timestep (the beginning or end of a note). 
This evaluation criterion is mostly invariant to the rate of temporal discretization. 


\subsection{Separated composition experiments}
\label{sec:exp_sep}

\begin{table*}[t]
\centering
\footnotesize
\begin{tabular}{lcccccccccccc}
	\toprule
    & \multicolumn{6}{c}{Negative log-likelihood} & \multicolumn{6}{c}{Accuracy}  \\[2pt]
    & \multicolumn{4}{c}{Single voice} & \multicolumn{2}{c}{Aggregate} & \multicolumn{4}{c}{Single voice} & \multicolumn{2}{c}{Aggregate} \\
     \cmidrule(lr){2-5} \cmidrule(lr){6-7} \cmidrule(lr){8-11} \cmidrule(lr){12-13}
    Model & P1 & P2 & TR & NO & POI & All & P1 & P2 & TR & NO & POI & All \\
    \midrule
Random & $4.36$ & $4.36$ & $4.49$ & $2.83$ & $16.04$ & $16.04$ & $.013$ & $.013$ & $.011$ & $.059$ & $.024$ & $.024$ \\
Unigram & $4.00$ & $3.77$ & $3.01$ & $2.50$ & $13.27$ & $11.53$ & $.020$ & $.022$ & $.057$ & $.061$ & $.040$ & $.369$ \\
Bigram & $4.91$ & $4.93$ & $4.15$ & $3.52$ & $17.50$ & $3.63$ & $.000$ & $.000$ & $.000$ & $.000$ & $.000$ & $.831$ \\
RNN Soloists & $4.92$ & $4.90$ & $3.59$ & $2.23$ & $15.64$ & $3.11$ & $.000$ & $.000$ & $.004$ & $.183$ & $.047$ & $.830$ \\
LSTM Soloists & $4.60$ & $4.30$ & $3.01$ & $1.91$ & $13.82$ & $2.70$ & $.014$ & $.008$ & $.125$ & $.246$ & $.098$ & $.838$ \\
LSTM Quartet & $3.87$ & $3.71$ & $2.45$ & $1.62$ & $11.65$ & $2.21$ & $.028$ & $.031$ & $.294$ & $.449$ & $.201$ & $.854$ \\
DeepBach~\cite{hadjeres2017deepbach} & $0.82$ & $1.01$ & $0.63$ & $0.83$ & $3.28$ & $0.75$ & $.781$ & $.729$ & $.784$ & $.748$ & $.761$ & $.943$ \\
    \bottomrule
\end{tabular}
\caption{Results for separated composition experiments. 
For each instrument, negative log-likelihood and accuracy are calculated at points of interest (POIs). We also calculate aggregate statistics at POIs and globally (All). While DeepBach~\cite{hadjeres2017deepbach} achieves the best statistical performance, it uses future context and hence is more expensive to sample from.}
\label{tab:sep}
\end{table*}


For separated composition, 
we evaluate the performance of several baselines and compare them to a cutting edge method. 
Our simplest baselines are 
unigram and additive-smoothed bigram distributions for each instrument. 
The predictions of such models are trivial; 
the unigram model always predicts ``no note'' and the bigram model always predicts ``last note''. 
The respective accuracy of these models, 
$37\%$ and $83\%$, 
reflect the proportion of the timesteps that are silent (unigram) or identical to the last timestep (bigram). 
However, 
if we evaluate these models only at POIs, 
their performance is substantially worse ($4\%$ and $0\%$).

We also measure performance of recurrent neural networks (RNNs) at modeling the voices independently. 
We train a separate RNN 
(either a basic RNN cell or an LSTM cell~\cite{hochreiter1997long}) 
on each voice to form our RNN~Soloists and LSTM~Soloists baselines. 
We compare these to LSTM~Quartet, 
a model consisting of a single LSTM that processes all four voices and outputs an independent softmax over each note category, giving the model full context of the composition in progress. 
All RNNs have $2$ layers and $256$ units, 
except for soloists which have $64$ units each, 
and we train them with $512$ steps of unrolling for backpropagation through time. 
We train all models to minimize NLL using the Adam optimizer~\cite{kingma2014adam} 
and employ early stopping based on the NLL of the validation set. 

While the DeepBach model~\cite{hadjeres2017deepbach} was designed for modeling the chorales of J.S. Bach, 
the four-voice structure of those chorales is shared by \nesmdb{}, 
making the model appropriate for evaluation in our setting. 
DeepBach embeds each timestep of the four-voice score 
and then processes these embeddings with a bidirectional LSTM to aggregate past and future musical context. 
For each voice, 
the activations of the bidirectional LSTM are concatenated with an embedding of all of the other voices, 
providing the model with a mechanism to alter its predictions for any voice in context of the others at that timestep. 
Finally, these merged representations are concatenated to an independent softmax for each of the four voices. 
Results for DeepBach and our baselines appear in \cref{tab:sep}. 


As expected, 
the performance of all models at POIs is worse than the global performance. 
DeepBach achieves substantially better performance at POIs than the other models, 
likely due to its bidirectional processing which allows the model to ``peek'' at future notes. 
The LSTM Quartet model is attractive because, 
unlike DeepBach, 
it permits efficient ancestral sampling. 
However, 
we observe qualitatively that samples from this model are musically unsatisfying. 
While the performance of the soloists is worse than the models which examine all voices, 
the superior performance of the LSTM Soloists to the RNN Soloists suggests that 
LSTMs may be beneficial in this context. 

We also experimented with artificially emphasizing POIs during training, 
however we found that resultant models produced unrealistically sporadic music. 
Based on this observation, 
we recommend that researchers who study \nesmdb{} always train models with unbiased emphasis, 
in order to effectively capture the semantics of the particular temporal discretization.

\subsection{Expressive performance experiments}
\label{sec:exp_expr}

\begin{table*}[t]
\centering
\footnotesize
\begin{tabular}{lcccccccccccccc}
	\toprule
    & \multicolumn{7}{c}{Negative log-likelihood} & \multicolumn{7}{c}{Accuracy}  \\[2pt]
    & \multicolumn{5}{c}{Single voice} & \multicolumn{2}{c}{Aggregate} & \multicolumn{5}{c}{Single voice} & \multicolumn{2}{c}{Aggregate} \\
     \cmidrule(lr){2-6} \cmidrule(lr){7-8} \cmidrule(lr){9-13} \cmidrule(lr){14-15}
    Model & $V_{\text{P1}}$ & $V_{\text{P2}}$ & $V_{\text{NO}}$ & $T_{\text{P1}}$ & $T_{\text{P2}}$ & POI & All & $V_{\text{P1}}$ & $V_{\text{P2}}$ & $V_{\text{NO}}$ & $T_{\text{P1}}$ & $T_{\text{P2}}$ & POI & All \\
    \midrule
Random & $2.77$ & $2.77$ & $2.77$ & $1.39$ & $1.39$ & $11.09$ & $11.09$ & $.062$ & $.062$ & $.062$ & $.250$ & $.250$ & $.138$ & $.138$ \\
Unigram & $2.87$ & $2.89$ & $3.04$ & $1.35$ & $1.33$ & $11.47$ & $9.65$ & $.020$ & $.022$ & $.061$ & $.006$ & $.004$ & $.023$ & $.309$ \\
Bigram & $2.82$ & $2.85$ & $2.78$ & $4.27$ & $4.27$ & $17.00$ & $4.57$ & $.000$ & $.000$ & $.000$ & $.000$ & $.000$ & $.000$ & $.741$ \\
MultiReg Note & $2.74$ & $2.72$ & $2.23$ & $1.27$ & $1.18$ & $10.13$ & $8.49$ & $.106$ & $.122$ & $.292$ & $.406$ & $.507$ & $.287$ & $.359$ \\
MultiReg Note+Auto & $2.58$ & $2.56$ & $2.04$ & $2.90$ & $2.48$ & $12.56$ & $4.32$ & $.073$ & $.100$ & $.345$ & $.071$ & $.096$ & $.137$ & $.752$ \\
LSTM Note & $2.68$ & $2.63$ & $2.09$ & $1.32$ & $1.21$ & $9.94$ & $8.28$ & $.115$ & $.134$ & $.305$ & $.456$ & $.532$ & $.308$ & $.365$ \\
LSTM Note+Auto & $1.93$ & $1.89$ & $1.99$ & $2.23$ & $1.89$ & $9.93$ & $3.42$ & $.305$ & $.321$ & $.386$ & $.241$ & $.432$ & $.337$ & $.774$ \\
    \bottomrule
\end{tabular}
\caption{Results for expressive performance experiments evaluated at points of interest (POI). Results are broken down by expression category (e.g. $V_{\text{NO}}$ is noise velocity, $T_{\text{P1}}$ is pulse 1 timbre) and aggregated at POIs and globally (All).}
\label{tab:perf}
\end{table*}

The expressive performance task consists of learning a mapping from a separated score to suitable expressive characteristics. 
Each timestep of a separated score in \nesmdb{} has 
note information (random variable $N$) for the four instrument voices.
An expressive score additionally has velocity ($V$) and timbre ($T$) information for P1, P2, and NO but not TR. 
We can express the distribution of performance characteristics given the composition as $P(V,~T \mid N)$. 
Some of our proposed solutions factorize this further into a conditional autoregressive formulation 
$\prod_{t=1}^{T} P(V_{t}, T_{t} \mid N, V_{\hat{t} < t}, T_{\hat{t} < t})$, 
where the model has explicit knowledge of its decisions for velocity and timbre at earlier timesteps. 

\begin{figure}[h]
\centering
\includegraphics[width=0.8\linewidth]{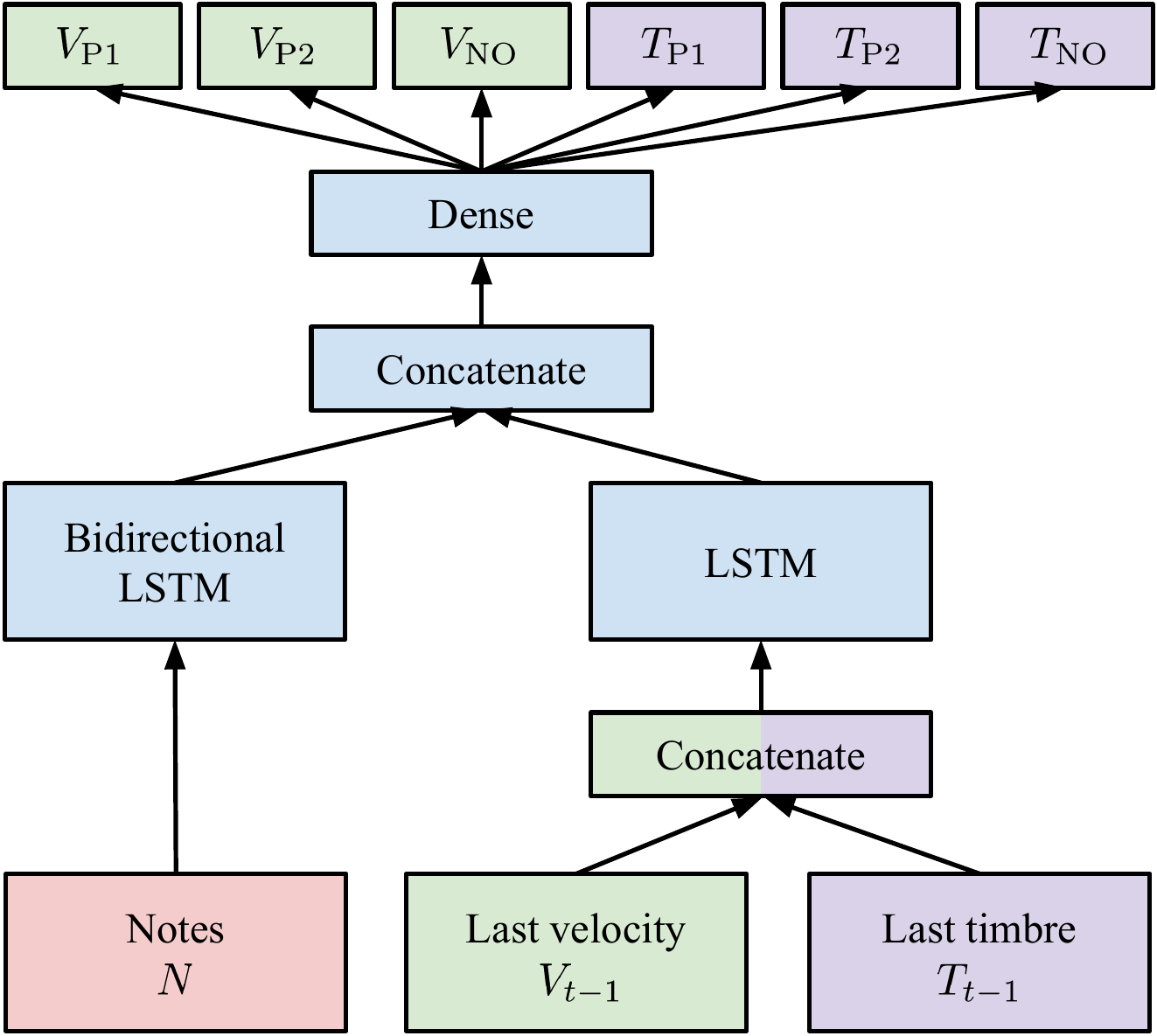}
\caption{LSTM Note+Auto expressive performance model that observes both the score and its prior output.}
\label{fig:perfmodel}
\end{figure}

Unlike for separated composition, 
there are no well-established baselines for multi-instrumental expressive performance, 
and thus we design several approaches. 
For the autoregressive formulation, 
our most-sophisticated model (\cref{fig:perfmodel}) uses a bidirectional LSTM to process the separated score, 
and a forward-directional LSTM for the autoregressive expressive characteristics. 
The representations from the composition and autoregressive modules are merged and processed by an additional dense layer before projecting to six softmaxes, one for each of 
$V_{\text{P1}}$, 
$V_{\text{P2}}$, 
$V_{\text{NO}}$, 
$T_{\text{P1}}$, 
$T_{\text{P2}}$, 
and 
$T_{\text{NO}}$. 
We compare this model (LSTM Note+Auto) to a version which removes the autoregressive module and only sees the separated score (LSTM Note). 

We also measure performance of simple multinomial regression baselines. 
The non-autoregressive baseline (MultiReg Note) maps the concatenation of 
$N_{\text{P1}}$, 
$N_{\text{P2}}$, 
$N_{\text{TR}}$, 
and 
$N_{\text{NO}}$ 
directly to the six categorical outputs representing velocity and timbre (no temporal context). 
An autoregressive version of this model (MultiReg Note+Auto) takes additional inputs consisting of the previous timestep for the six velocity and timbre categories. 
Additionally, 
we show results for simple baselines (per-category unigram and bigram distributions) which do not consider $N$. 
Because the noise timbre field $T_{\text{NO}}$ is so rarely used (less than $0.2\%$ of all timesteps), we exclude it from our quantitative evaluation. 
Results are shown in \cref{tab:perf}. 

Similarly to the musical notes in the separated composition task (\cref{sec:exp_sep}), 
the high rate of \nesmdb{} results in substantial redundancy across timesteps. 
Averaged across all velocity and timbre categories, 
any of these categories at a given timestep has a $74\%$ chance of having the same value as the previous timestep. 

The performance of the LSTM Note model is comparable to that of the LSTM Note+Auto model at POIs, 
however the global performance of the LSTM Note+Auto model is substantially better. 
Intuitively, 
this suggests that the score is useful for knowing \emph{when} to change, 
while the past velocity and timbre values are useful for knowing \emph{what} value to output next. 
Interestingly, 
the MultiReg Note model has better performance at POIs than the MultiReg Note+Auto model. 
The latter overfit more quickly 
which may explain its inferior performance despite the fact that it sees strictly more information than the note-only model.


\subsection{Blended composition experiments}
\label{sec:exp_blend}

\begin{table}[t]
\centering
\footnotesize
\begin{tabular}{lccccc}
	\toprule
    Model& \clap{\nesmdb} & PM & NH & MD & BC \\
    \midrule
Random & $61.00$ & $61.00$ & $61.00$ & $61.00$ & $61.00$ \\
Note 1-gram~\cite{boulanger2012modeling} & $8.71$ & $11.05$ & $10.25$ & $11.51$ & $11.06$ \\
Chord 1-gram~\cite{boulanger2012modeling} & $8.76$ & $27.64$ & $5.94$ & $19.03$ & $12.22$ \\
GMM~\cite{boulanger2012modeling} & $12.86$ & $15.84$ & $7.87$ & $12.20$ & $11.90$ \\
NADE~\cite{boulanger2012modeling} & $8.53$ & $10.28$ & $5.48$ & $10.06$ & $7.19$ \\
RNN~\cite{boulanger2012modeling} & $3.04$ & $8.37$ & $4.46$ & $8.13$ & $8.71$ \\
RNN-NADE~\cite{boulanger2012modeling} & $2.62$ & $7.48$ & $2.91$ & $6.74$ & $5.83$ \\
LSTM & $2.54$ & $8.31$ & $3.49$ & $6.35$ & $8.72$ \\
LSTM-NADE~\cite{johnson2017generating} & $2.48$ & $7.36$ & $2.02$ & $5.02$ & $6.00$ \\
    \bottomrule
\end{tabular}
\caption{Negative log-likelihoods for various models on the blended score format (\cref{fig:rep_blend}, \cref{eq:blend}) of \nesmdb{}. We also show results for Piano-midi.de (PM), Nottingham (NH), MuseData (MD), and the chorales of J.S.~Bach (BC).
}
\label{tab:blended}
\end{table}

In \cref{tab:blended}, 
we report the performance of several models on the blended composition task (\cref{eq:blend}). 
In \nesmdb, 
blended scores consist of $88$ possible notes with a maximum of three simultaneous voices (noise generator is discarded). 
This task, 
standardized in~\cite{boulanger2012modeling}, 
does not preserve the voicing of the score, 
and thus it is not immediately useful for generating NES music. 
Nevertheless, 
modeling blended scores of polyphonic music 
has become a standard benchmark for sequential models~\cite{chung2014empirical,jozefowicz2015empirical}, 
and \nesmdb{} may be useful as a larger dataset in the same format. 

In general, 
models assign higher likelihood to \nesmdb{} than the four other datasets after training. 
As with our other two tasks, 
this is likely due to the fact that \nesmdb{} is sampled at a higher temporal rate, 
and thus the 
average deviation across timesteps is lower. 
Due to its large size, a benefit of examining \nesmdb{} in this context is that sequential models tend to take longer to overfit the dataset than they do for the other four. 
We note that our implementations of these models may deviate slightly from those of the original authors, 
though our models achieve comparable results to those reported in~\cite{boulanger2012modeling,johnson2017generating} when trained on the original datasets.


\section{Related work}

There are several popular datasets commonly used in statistical music composition. 
A dataset consisting of the entirety of J.S. Bach's four-voice chorales has been extensively studied under the lenses of algorithmic composition and reharmonization~\cite{hild1992harmonet,allan2005harmonising,boulanger2012modeling,hadjeres2017deepbach}. 
Like \nesmdb, this dataset has a fixed number of voices and can be represented as a separated score (\cref{fig:rep_sep}), 
however it is small in size ($389$ chorales) and lacks expressive information. 
Another popular dataset is Piano-midi.de, a corpus of classical piano from various composers~\cite{poliner2006discriminative}. 
This dataset has expressive timing and dynamics information but has heterogeneous time periods and only features solo piano music. 
Alongside Bach's chorales and the Piano-midi.de dataset, 
Boulanger-Lewandowski~et~al.~(2012) standardized the Nottingham collection of folk tunes and MuseData library of orchestral and piano classical music into blended score format (\cref{fig:rep_blend}). 

Several other symbolic datasets exist containing both compositional and expressive characteristics. 
The Magaloff Corpus~\cite{flossmann2010magaloff} consists of Disklavier recordings of a professional pianist playing the entirety of Chopin's solo piano works. 
The Lakh MIDI dataset~\cite{raffel2016learning} is the largest corpus of symbolic music assembled to date with nearly $200$k songs. 
While substantially larger than \nesmdb{}, 
the dataset has unconstrained polyphony, 
inconsistent expressive characteristics, 
and encompasses a wide variety of genres, instruments and time periods.
Another paper trains neural networks on transcriptions of video game music~\cite{fabius2015variational}, 
though their dataset only includes a handful of songs.

\subsection{Statistical composition}


While most of the early research in algorithmic music composition focused on expert systems~\cite{nierhaus2009algorithmic}, 
statistical approaches have since become the predominant approach. 
Mozer (1994) trained RNNs on monophonic melodies using a formulation similar to \cref{eq:mono}, finding the composed results to compare favorably to those from a trigram model. 
Others have also explored monophonic melody generation with RNNs~\cite{eck2002finding,paiement2009probabilistic}. 
Boulanger-Lewandowski et al. (2012) standardize the polyphonic prediction task for blended scores (\cref{eq:blend}), 
measuring performance of a multitude of classical baselines against RNNs~\cite{rumelhart1986learning}, restricted Boltzmann machines~\cite{smolensky1986information}, and NADEs~\cite{larochelle2011neural} on polyphonic music datasets. 
Several papers~\cite{chung2014empirical,vohra2015modeling,johnson2017generating} directly compare to their results.
Statistical models of music have also been employed as symbolic priors
to assist music transcription algorithms~\cite{cemgil2004bayesian,nam2011classification,boulanger2012modeling}. 

Progressing towards models that \emph{assist} humans in composition, 
many researchers study models to create new harmonizations for existing musical material. 
Allan and Williams (2005) train HMMs to create new harmonizations for Bach chorales~\cite{allan2005harmonising}. 
Hadjeres et al. (2017) train a bidirectional RNN model to consider past and future temporal context (\cref{eq:sep})~\cite{hadjeres2017deepbach}. 
Along with~\cite{sakellariou2015maximum,huang2016counterpoint}, they advocate for the usage of Gibbs sampling to generate music from complex graphical models.

\subsection{Statistical performance}

Musicians perform music expressively by interpreting a performance with appropriate dynamics, timing and articulation. 
Computational models of expressive music performance seek to automatically assign such attributes to a score~\cite{widmer2004computational}. 
We point to several extensive surveys for information about the long history of 
rule-based systems~\cite{widmer2004computational,goebl2008sense,delgado2011state,kirke2013overview}. 

Several statistical models of expressive performance have also been proposed.
Raphael (2010) learns a graphical model that automates an accompanying orchestra for a soloist, operating on acoustic features rather than symbolic~\cite{raphael2010music}. 
Flossmann et al. (2013) build a system to control velocity, articulation and timing of piano performances by learning a graphical model from a large symbolic corpus of human performances~\cite{flossmann2013expressive}. 
Xia et al. (2015) model the expressive timing and dynamics of piano duet performances using spectral methods~\cite{xia2015spectral}. 
Two end-to-end systems attempt to jointly learn the semantics of composition and expressive performance using RNNs~\cite{simon2017performance,mao2018deepj}. 
Malik and Ek (2017) train an RNN to generate velocity information given a musical score~\cite{malik2017neural}. 
These approaches differ from our own in that they focus on piano performances rather than multi-instrumental music.

\section{Conclusion}

The NES Music Database is a large corpus for examining multi-instrumental polyphonic composition and expressive performance generation. 
Compared to existing datasets, 
\nesmdb{} allows for examination of the ``full pipeline'' of music composition and performance. 
We parse the machine code of NES music into familiar formats (e.g. MIDI), 
eliminating the need for researchers to understand low-level details of the game system. 
We also provide an open-source tool which converts between the simpler formats and machine code, 
allowing researchers to audition their generated results as waveforms rendered by the NES.
We hope that this dataset will facilitate a new paradigm of research on music generation---one that emphasizes the importance of expressive performance. 
To this end, we establish several baselines with reproducible evaluation methodology to encourage further investigation.



\section{Acknowledgements}

We would like to thank Louis Pisha for invaluable advice on the technical details of this project. 
Additionally, we would like to thank 
Nicolas Boulanger-Lewandowski, 
Eunjeong Stella Koh, 
Steven Merity, 
Miller Puckette, 
and Cheng-i Wang for helpful conversations throughout this work. 
This work was supported by UC San Diego's Chancellor’s Research Excellence Scholarship program. 
GPUs used in this research were donated by NVIDIA. 


\bibliography{ISMIRtemplate}

\end{document}